%% file: main.tex
\documentclass{article}

\usepackage{arxiv}

\usepackage[utf8]{inputenc} 
\usepackage[T1]{fontenc}    
\usepackage{hyperref}       
\usepackage{url}            
\usepackage{booktabs}       
\usepackage{amsfonts}       
\usepackage{nicefrac}       
\usepackage{microtype}      
\usepackage{lipsum}

\usepackage[utf8]{inputenc}
\usepackage{hyperref,comment}
\usepackage{amsmath}
\usepackage{amsthm}
\usepackage{amsfonts}
\usepackage[usenames,dvipsnames]{color}
\usepackage{graphicx}
\definecolor{pink}{rgb}{0.858, 0.188, 0.478}
\usepackage[round]{natbib}

\newcommand{\N}{\mathbb{N}}

\title{Smarter Parking: Using AI to Identify Parking Inefficiencies in Vancouver}
\date{}

\author{
  Devon Graham \\
  Department of Computer Science\\
  University of British Columbia\\
  Vancouver, BC \\
  \texttt{drgraham@cs.ubc.ca} \\
  \And 
  Satish Kumar Sarraf \\
  Department of Electrical and Computer Engineering\\
  University of British Columbia\\
  Vancouver, BC \\
  \texttt{ssatish@ece.ubc.ca} \\
  \And 
  Taylor Lundy \\
  Department of Computer Science\\
  University of British Columbia\\
  Vancouver, BC \\
  \texttt{tlundy@cs.ubc.ca} \\
  \And 
  Ali MohammadMehr \\
  Department of Computer Science\\
  University of British Columbia\\
  Vancouver, BC \\
  \texttt{alimm@cs.ubc.ca} \\
   \And
  Sara Uppal \\
  Department of Electrical and Computer Engineering\\
  University of British Columbia\\
  Vancouver, BC \\
  \texttt{sarau@ece.ubc.ca} \\
  \And 
  Tae Yoon Lee \\
  Department of Statistics\\
  University of British Columbia\\
  Vancouver, BC \\
  \texttt{taeyoon.lee@alumni.ubc.ca} \\
  \And 
  Hedayat Zarkoob \\
  Department of Computer Science\\
  University of British Columbia\\
  Vancouver, BC \\
  \texttt{hzarkoob@cs.ubc.ca} \\
  \And
   Scott Duke Kominers \\
  Harvard Business School \\
  Boston, MA \\
  \texttt{kominers@fas.harvard.edu}
  \And
  Kevin Leyton-Brown \\
  Department of Computer Science\\
  University of British Columbia\\
  Vancouver, BC \\
  \texttt{kevinlb@cs.ubc.ca}
  }

\begin{document}
\date{}
\maketitle

\begin{abstract}
    \label{abstract}
    \input{abstract}

\end{abstract}

\keywords{Parking \and Street Parking \and Deep Neural Networks \and Occupancy Prediction \and Parking Simulation}

\section{Introduction}
\label{intro}
\input{intro}

\section{Literature Review}
\label{lit-review}
\input{lit-review}

\section{Methods}
\label{methods}
\input{methods}

\section{Data}
\label{data}

\input{data}

\section{Results}
\label{result}
\input{results}

\section{Conclusion}
\label{conclusion}
\input{conclusion}

  \bibliographystyle{named}
  \bibliography{main}

\end{document}

%% file: abstract.tex


On-street parking is convenient, but has many disadvantages: on-street spots come at the expense of other road uses such as traffic lanes, transit lanes, bike lanes, or parklets; drivers looking for parking contribute substantially to traffic congestion and hence to greenhouse gas emissions; safety is reduced both due to the fact that drivers looking for spots are more distracted than other road users and that people exiting parked cars pose a risk to cyclists. These social costs may not be worth paying when off-street parking lots are nearby and have surplus capacity. To see where this might be true in downtown Vancouver, we used artificial intelligence techniques to estimate the amount of time it would take drivers to both park on and off street for destinations throughout the city. For on-street parking, we developed (1) a deep-learning model of block-by-block parking availability based on data from parking meters and audits and (2) a computational simulation of drivers searching for an on-street spot. For off-street parking, we developed a computational simulation of the time it would take drivers drive from their original destination to the nearest city-owned off-street lot and then to queue for a spot based on traffic and lot occupancy data. Finally, in both cases we also computed the time it would take the driver to walk from their parking spot to their original destination.
We compared these time estimates for destinations in each block of Vancouver's downtown core and each hour of the day. We found many areas where off street would actually save drivers time over searching the streets for a spot, and many more where the time cost for parking off street was small. The identification of such areas provides an opportunity for the city to repurpose valuable curbside space for community-friendly uses more in line with its transportation goals.

%% file: intro.tex
In many urban areas, trying to find curbside parking can be an unpleasant experience. In peak hours, available spots are rare and drivers can cruise from block to block without finding one. Even having found a spot, drivers may then need to walk a significant distance back to their destination. This takes time and clogs up city streets with unnecessary congestion. In fact, in a well-known study, Shoup found that on average 30\% of traffic on city streets can be the result of such cruising~\citep{shoup2006cruising}. This excessive congestion has many disadvantages. Drivers searching for parking are distracted, reducing safety for pedestrians and cyclists~\citep{edquist2011effects}. With more cars on the road, travel times are increased for everyone, including public transit vehicles. More slow-moving cars on roads means increased emissions~\citep{rouphail2001vehicle}. Thus, it is important for modern cities to manage their parking resources efficiently.  

As an alternative to curbside parking, drivers have the option to park in off-street lots (by which we refer equally to parkades, parking garages, and ground-level lots). Such lots often offer many more available spots than on-street parking. However, parking off-street has its own drawbacks. Drivers need to first find the lot, find a free spot, pay, and then walk back to their destination. It is therefore not immediately clear whether most drivers save time by seeking on- or off-street parking. 

Previous studies have examined the trade-off between searching for curbside parking and parking in the nearest lot, most through the lens of examining pricing schemes~\citep{RePEc:eee:trapol:v:25:y:2013:i:c:p:222-232,shoup2006cruising}. In particular, Shoup looked at how drivers might think about on- and off-street parking by assuming that curbside parking is free but rare while off-street spots are always available but  expensive~\citep{shoup2006cruising}. He observed that if the time it takes for drivers to cruise for free curbside parking exceeds a certain threshold, they will prefer paying for off-street parking to save time. However, price is not the only factor affecting drivers' decisions~\citep{yan2019effectiveness}. Furthermore, drivers trade time for money differently depending on their situations. If a driver has an important meeting with a client, she will gladly pay more for parking to avoid being late and risking losing the client; if a driver holds an expensive concert ticket, paying a high price for parking is not as bad as missing part of the show. 


Our work approaches the problem of curbside vs.\ lot parking from another angle: we consider only the difference in time between the two alternatives, and think of price as a tool that can be used to change driver behaviour. The off-street lots we consider are all owned by the City of Vancouver---as, of course, are the on-street spots---meaning that adjusting prices is within the power of the City's decision makers. 
We show how cutting-edge AI tools can be leveraged along with real-world data to build realistic models of on- and off-street parking procedures. We show that despite the costs of off-street parking mentioned above, drivers can sometimes save time by parking in off-street lots rather than cruising in search of on-street parking. 

In order to simulate cruising for parking, we predict curbside parking occupancy from observed payment data.
Various methods exist for making such predictions \citep{rajabioun2015street,tamrazian2015my,yang2017turning,yang2019deep}, but none of these methods works with the sort of highly incomplete data we encountered in practice. Aggregated payment data is only weakly predictive of actual occupancy rates~\citep{RePEc:eee:trapol:v:25:y:2013:i:c:p:222-232}: illegally parked cars still consume a spot, and paid drivers can leave both before and after their meter expires. While curbside payments are performed via on-street meters that accept credit cards, cash, and app payments, we only had access to app payments. We were able to calibrate these noisy observations with a very limited number of occupancy checks performed by a parking inspector. We show that in this setting it is nevertheless possible to build accurate models using deep neural networks.

The remainder of this paper begins with a brief literature review in Section~\ref{lit-review}, contrasting approaches taken in the literature with our own methods. Then, we describe the methods we developed to simulate both on- and off-street parking in Section~\ref{methods}. Section~\ref{data} describes the data used in our simulations and Section~\ref{result} analyzes the results. Finally, we conclude and discuss some future directions in Section~\ref{conclusion}.

%% file: lit-review.tex
Others have previously considered methods for simulating the parking search procedure, as well as comparing on- and off-street parking, and predicting occupancy from payment data. We review some relevant work below, highlighting methodological differences in the approaches proposed in this paper.

\subsection{On-Street Parking Simulators}
 
 \citet{mannini2017street} performed an influential study of on-street parking time estimation, computing on-street parking search time by using Floating Car Data obtained from probe vehicles.  \citet{belloche2015street} used survey-based data to compute parking search time based on an off-street parking search time model proposed by \cite{axhausen1994effectiveness}. 
 Most related to our own work, \citet{dowling2019modeling} viewed streets as a network, representing streets and intersections as edges and nodes in a graph. Then, using the transaction data to estimate true occupancy,~\cite{dowling2019modeling} calculated various network-level measures such as congestion. Our work uses transaction data to estimate true occupancy using a deep learning model and estimates the parking search time with a simple search algorithm using Monte-Carlo simulation.

\subsection{Off-Street Parking Simulators}
While there exist agent-based choice models for parking in general \citep{vuurstaek2018first,benenson2008parkagent,waraich2012agent,bischoff2017integrating} and survey-based studies for off-street parking search time \citep{axhausen1994effectiveness,teng2002parking}, literature on explicitly modeling parking in an off-street parking lot is quite limited. We propose a simple simulation-based approach to compute the time it takes for a car to park in an off-street parking lot.

\subsection{On-Street vs. Off-Street Parking}
There has been active research in parking policy and regulation for both on-street parking \citep{marshall2008reassessing,biswas2017effects} and off-street parking \citep{barter2010off}. On-street parking is evaluated on a combination of three areas: safety for drivers, cyclists, and pedestrians; road congestion; and economic activity \citep{marshall2008reassessing}. Off-street parking lots serve as extra supply to accommodate excess on-street parking demand. Studies have suggested using pricing to manage the markets for on-street and off-street parking individually~\citep{RePEc:eee:trapol:v:25:y:2013:i:c:p:222-232}. However, these studies do not offer a collective assessment of on-street and off-street parking. \cite{inci2015garage} considers the relationship between on-street parking and off-street parking by examining price. Our own work uses time as a common metric to compare on- and off-street parking and identify destinations for which it takes less time to park off-street than on-street.

\subsection{Parking Occupancy Prediction from Transaction Data}
The most straightforward method of estimating parking occupancy from transaction data is simply to assume that every driver leaves exactly once her paid session expires and that all drivers pay for parking. \cite{fiez2017data} and \cite{dowling2019modeling} estimate parking occupancy in this way at a block-face level over the course of an hour. They acknowledge the limitations of this method, noting particularly that drivers tend to leave before their paid parking time, leading to over-estimation of occupancy. In recent years, statistical and machine learning models have been developed to deal with such issues and predict parking occupancy more accurately using additional data. These include multivariate spatial-temporal models \citep{rajabioun2015street}, unsupervised clustering ($k$-means and $k$-nearest neighbours) \citep{tamrazian2015my}, probabilistic model \citep{yang2017turning}, and most notably a series of deep learning approaches \citep{yang2019deep,Zheng2015ParkingAP,Kepaptsoglou2014ExploitingNS}.
%
%
These deep learning approaches rely heavily on the availability of actual occupancy data from sensors and surveys, which indeed is becoming increasingly available. Both \cite{yang2019deep} and \cite{Kepaptsoglou2014ExploitingNS} leveraged a model called recurrent neural nets to predict occupancy rates from complete transaction data. Another approach by \cite{Zheng2015ParkingAP} used the data from SFPark (approximately 2 million block-level samples) to construct a deep neural network, comparing performance both to regression trees and support vector machines. In contrast to all of this work, our training data is sparse in two senses. First, our transaction data is not complete: we have access only to a single mode of payment, which comprises about 60\% of total meter transactions. Second, the observations used for our ground truth occupancy levels are based on human surveys rather than sensors, giving us considerably less data than needed by the approaches just mentioned; furthermore, this data is not uniformly sampled across different times of day. This paper proposes an alternate deep learning approach for estimating and predicting parking occupancy by combining incomplete transaction data from parking meters with limited in-person counts.

%% file: methods.tex
\subsection{On-Street}
We construct a simulation of the time it takes to park on-street by modeling the behaviour of a driver searching for a spot under the assumption that the driver starts at his/her destination. To generate accurate search times, we first need predictions of the availability of parking at each individual block of the city, for every hour of the day. To generate these predictions we use a deep neural network trained to predict occupancy from payment data. In this section, we first give some brief background on feed-forward neural networks. We then describe the specific network we construct to solve this problem, and finally, we describe how we use these predictions to simulate search times for on-street parking.

\subsubsection{Neural Networks}\label{sec:nn}
Neural networks are a broad class of functions that map inputs to outputs in a complex fashion. These networks are built from a sequence of $k \in \N$ ``layers'', where the output of each layer is fed as the input to the next. Each layer consists of a non-linear function $\sigma$ applied to a linear transformation of its input. The linear transformation in each layer $i$ is
parameterized by a weight matrix, $W_i$, and a bias vector, $b_i$. Thus, a neural network is just a function of the form 
\begin{equation}
f(x) = W_k \; \sigma \bigg( W_{k-1} \; \sigma \bigg( \dots \; \sigma(W_2 \; \sigma( W_1 x + b_1 ) + b_2) \dots  \bigg) + b_{k-1} \bigg) + b_k, 
\label{eq1}
\end{equation}
where $x$ is the input. Neural networks can approximate any function to arbitrary precision, if the matrices $W$ are chosen to be large enough. Given inputs $x$ and target values $y$, training the network means finding values of the parameters $W$ and $b$ for each layer such that the network's output, $f(x)$ is close to the true, known output $y$.

\subsubsection{Occupancy Prediction Network}
We assume that we are given a partial record of payments made at parking meters in each block face in the city. Inferring occupancy from payment information is not as straightforward as it may seem: drivers may leave before their time is up, in which case payment data overestimates occupancy; alternately, they may overstay or even park without paying at all, in which case payment data will underestimate occupancy. Our goal is to predict the true occupancy of a block as a function of observed payment data for that block. 

\begin{figure}
    \centering
    \includegraphics[width=0.7\columnwidth]{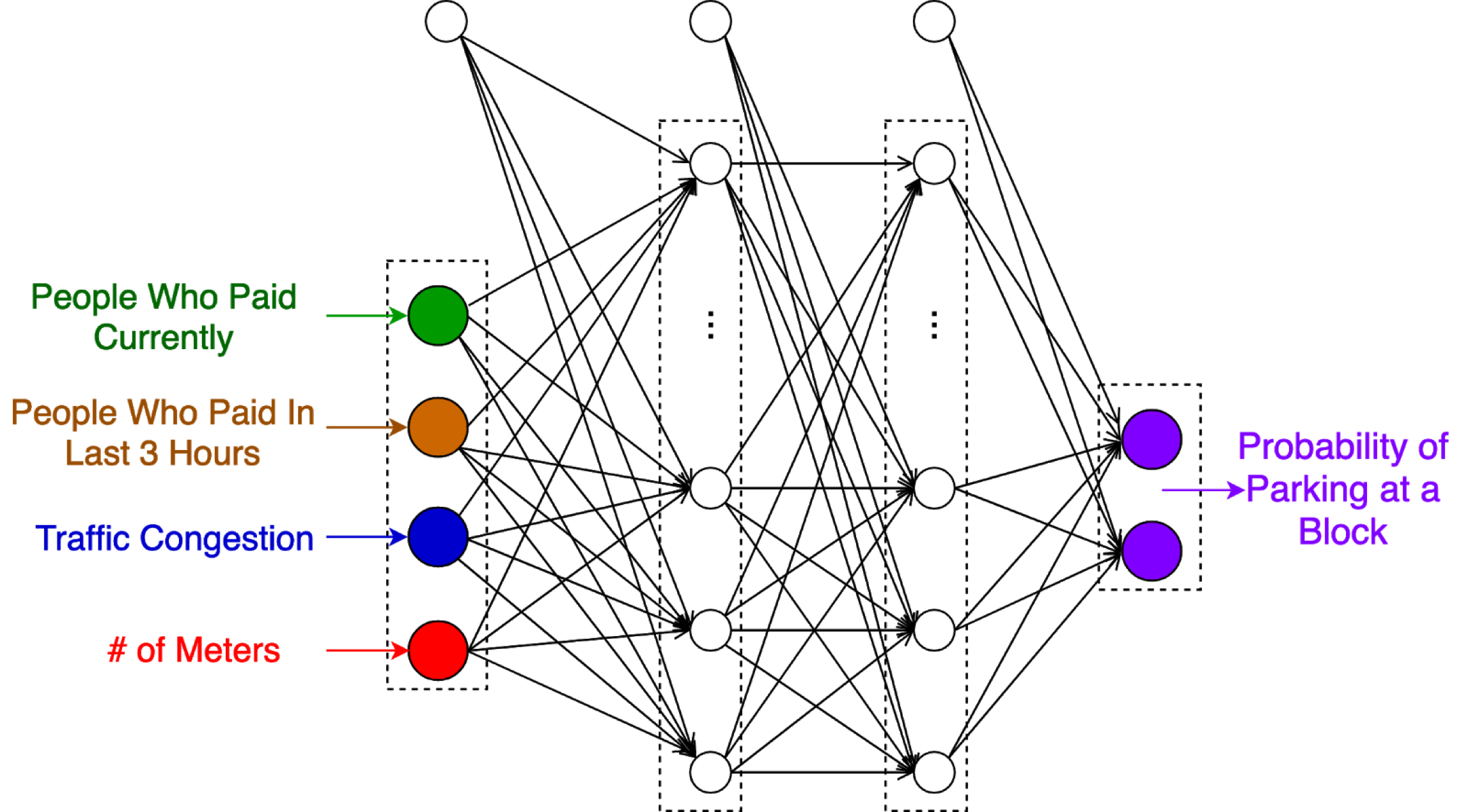}
    \caption{\textit{Neural network design used to compute the probability of parking at a block.}}
    \label{fig:my_label}
\end{figure}
For each block face and time, our model is given two input values which are derived from the payment data. We augment this with two input values that do not directly depend on the payment data, but relate to the block face in question. The input values are represented as a vector of length 4 consisting of the following values:
\begin{enumerate}
\item The total number of paid parking sessions in the block at the given time;
\item The total number of paid parking sessions in the block over the 3 hour period before the given time (``popularity'')
\item The length of the block in meters; and 
\item The predicted time to drive the length of the block divided by the length of the block (``congestion'').
\end{enumerate}


For each block and time, the neural network outputs a vector of length two, whose elements are forced to sum to 1 and are thus interpretable as probabilities. To train the network, we minimize the cross-entropy loss between these probabilities and the true, observed occupancy state of the block. In this way, the network learns to approximate the true probability of parking being available in a block at a given time, as a function of the given data. 
Using the notation of Section \ref{sec:nn}, our network consists of an input layer, two hidden layers and an output layer (thus $k =3$; the input layer does not count as a layer). The connections between these layers are weight matrices of sizes $4 \times 30$, $30 \times 30$, and $30 \times 2$ respectively. Including  bias vectors, this fully-connected neural network is described by $1,142$ parameters. 
As a non-linear function we use ReLU: $\sigma(x) = \max(0, x)$. We train the network on ten different random training and validation splits, with the validation set always constituting 20\% of the total data, and iterating over the training data 150 times for each split. 

Once the network is trained, we predict the existence of an empty meter at a given time (and thus whether a driver can park) in any block by simply feeding the model the payment and block-level features. Since the model has been trained to reproduce the probability that there will be an available spot, we can use this output directly in our simulation. To evaluate our network's performance, we compare its results to a logistic regression model as baseline that is given the same input variables our network is given. We evaluate both the network and the regression in terms of accuracy (with threshold $0.5$) and cross-entropy.





\subsubsection{On-street Search Simulator}
A key portion of our simulator models a driver's search for on-street parking, based on noisy observations of traffic meter usage. We simulate search time as the sum of three factors: the minimum search time, the estimated time required to drive on nearby streets while searching, and the estimated time required to walk back to the destination. 

More specifically, we constructed a representation of downtown Vancouver using OSMnx \citep{OSMnx}, a Python package for modeling road networks. The city is modeled as a graph structure:  intersections correspond to ``nodes'' and street segments correspond to ``edges'' connecting pairs of nodes. We assume that drivers begin searching for parking in the block that contains their destination.
Using the probabilities predicted by the neural network, we model the on-street parking search process for each block face. During the search process, for any edge the vehicle traverses, we say it can park on that block with the probability predicted by the neural network. If it parks, we end the search process.
We assume that on-street parking requires some minimum amount of time, $t_{\min}$; this corresponds to the time required for parallel parking and paying at the meter. We set $t_{\min}=210$ seconds, following~\citet{shoup2006cruising}. 
If the vehicle fails to park, it moves to the next edge and tries again. Once the vehicle is parked, the driver walks from the parking spot to the original destination; we assume both locations are in the middles of their respective blocks. 
We obtained driving and walking times from the Google Maps Directions API, assuming that walking times remain constant throughout the day. For a single vehicle, let $d_1, d_2, ..., d_n$ be the driving times for the blocks the vehicle traverses, and $w_1, w_2, ..., w_m$ be the walking times for blocks the 
driver traverses as a pedestrian.
Then the total on-street search time, $T$, is modeled as 
\begin{equation}
   T = t_{\min} +  \bigg( \frac{d_{1}}{2} + \sum_{i=2}^{n} d_{i} - \frac{d_{n}}{2} \bigg) + \bigg( \frac{w_{1}}{2} + \sum_{j=2}^{m} w_{j} - \frac{w_{m}}{2} \bigg).
\end{equation}
With each block face and for each time of the day, we use the simulator to produce many samples of the value $T$, and then average to estimate mean search time. 

A key consideration in simulating driver search is reasoning about how drivers choose which block to search next. We model this as a random process, where the probability of the driver choosing a block, $i$, depends on four quantities: 
\begin{enumerate}
    \item $D_i$, the distance between block $i$ and the driver's destination (drivers are more likely to turn back towards the destination than away, making blocks further away less likely to be selected);
    \item $N_i$, the number of times the driver has already checked block $i$ during their search (each repetition makes a block less likely to be checked);
    \item $E_i$, the time that has elapsed since the driver last checked this block (the more recent, the less likely the block is to be rechecked); and 
    \item $P_i$, the probability of parking being available on block $i$, as predicted by the neural network model (blocks with more available parking are more likely to be visited). Note that this only incorporates the parking probabilities adjacent to the driver's current intersection. This is because we assume that drivers do not have accurate predictions about parking probabilities across the whole city, but that drivers can visually inspect blocks adjacent to the intersection to obtain about as much information as given by the predictive model.
\end{enumerate}
We form a linear combination of these quantities, 
\begin{align*}
    Z_i = w_D D_i + w_N N_i + w_E E_i + w_P \frac{1}{P_i},
\end{align*}
where we hand tuned the $w$ parameters to produce results we considered realistic, setting them to $w_D = -1, w_N=-15, w_E=15, w_P = -1$. 

Observe that because the quantity $Z_i$ could be greater than 1, or even negative, we cannot interpret it as a probability. Instead, we interpret it as the log-odds of choosing this block, and convert it to a probability with the softmax function
\begin{align*}
    \Pr[\text{choose Block i}] = \frac{e^{Z_i}}{\sum_j e^{Z_j}}. 
\end{align*}
This produces a realistic approximation of a driver looking for parking since it (i) discourages drivers from travelling too far from their destinations; (ii) prevents drivers from re-checking the same block over and over; and (iii) incorporates knowledge of occupancy levels on the blocks the driver could reasonably see at the time of their decision. 
\subsection{Off-Street}

We estimate the time it takes to park in the closest off-street lot for each block face in the city. To do so, we compute the expected times for a driver (1) to drive from a given destination to that lot; to search for a spot in the lot and park; (3) to walk back to their destination. To compute (1) and (3)---taking into account likely traffic at a given time of day---we appeal to historical traffic data via the Google Maps Directions API. To estimate (2), the time spent searching in the lot, we use a Monte Carlo simulation, described next.

\subsubsection{Time Spent Searching in Lots}

Our simulation of the time spent in a lot is based on entry and exit data collected from several city-owned lots. For simplification, we assume that all lots are one-dimensional manifolds: i.e., that they have a single entrance and exit with no branching paths. We assume parking in a lot requires some minimum amount of time $t'_{\min}$. In practice we set $t'_{\min} = 60$ seconds. (This may be an overestimate, but if so this only serves to help our argument that lots can often be better for drivers than on-street parking.) We model the time a driver spends in the lot as influenced not only by the number of vehicles currently in the lot, but also by the number of other vehicles entering and exiting the lot at the same time. 


We consider a small unit of time (in practice, 60 seconds) and model the number of cars to arrive and to depart in that unit of time as Poisson distributed random variables with means $\lambda_a$ and $\lambda_d$, respectively. The parameters $\lambda_a$ and $\lambda_d$ are determined from the observed entry and exit data for each lot and each hour of the day and for each day of the week, averaged over 12 weeks. We keep track of the state of the lot by updating each spot in each unit of time. When it is time for a vehicle to depart, we select that vehicle uniformly at random among the lot's occupied spots. 


We assume that vehicles park in the same order that they arrive, and that each one parks in the first available spot it finds. We assume that drivers will wait $30$ seconds for a departing car and then take that spot if necessary. Vehicles need to wait for those in front of them to park. The $k^{\text{th}}$ vehicle to arrive may wait for up to $k$ departing vehicles to vacate stalls, or may wait for none, so we assume that on average it waits for half of them. While driving in the lot, we use a constant time $t_{1}$ as the time it takes to drive past a single parking stall. 

Because the minimum time $t'_{\min}$ includes paying for parking, and unlike pulling into spots, not all vehicles can always pay at the same time, we assume that the $k^{\text{th}}$ vehicle must also wait 1/2 of the minimum time for the vehicle in front of it, 1/4 of the minimum time of the second vehicle in front of it, 1/8 of the minimum time of the third vehicle, etc. 

Then, if $N_a \sim_d Poisson(\lambda_a)$ vehicles arrive and $N_d \sim_d Poisson(\lambda_d)$ vehicles depart in this small period of time, the $k^{\text{th}}$ vehicle to arrive waits for $T_k$ seconds, where
\begin{equation}
     T_k = t'_{\min} + S_k t_1 + \frac{\min(k, N_d)}{2} t_{\text{wait}} + \sum_{i=1}^{k-1} \frac{1}{2^i} t_{\min}
     \label{eq3}
\end{equation}
seconds, where $S_k$ is the spot that the $k^{\text{th}}$ driver parks in according to the simulation, updated based on $N_a$. For each $k = 1, ..., N_a$, we consider $T_k$ as a single Monte Carlo sample, repeat the process 20 times, and average over all samples to determine the total average search time for each lot and each time of day. We used a value of $t_1 = 0.54$ seconds, based on data recording parking activity in a large local lot, averaging the time it takes to drive through the whole lot (while following the posted speed limit) over the lot's total number of stalls.

%% file: data.tex
\subsection{Geographic Data}

The city of Vancouver offers many publicly available data sets in its \emph{Open Data Catalogue} ({{\url{https://data.vancouver.ca/datacatalogue}}}). We made use of two of these: \emph{Parking Meter Data} contains the location, rates and time limits for approximately 10,000 parking meters in the city, and \emph{City Streets} associates each meter in the Parking Meter dataset with its corresponding block face.

\subsection{Occupancy Surveys}

To train our  neural network for predicting on-street parking spot occupancy, we used (non-publicly available) data from occupancy surveys done by the City of Vancouver on each of the meters on 372 block faces in Vancouver's downtown area. The survey data consisted of timestamped, meter-level surveys. However, since the meters in each block face were surveyed at nearly the same time, we combined all surveys of the meters in a block face that occurred in the same 30-minute period to obtain a single snapshot of the block face. Overall, each block face was checked on average 8 times, with 4 checks in the morning and 4 checks in the afternoon; after combining meter-level data, we had a total of 3084 block-level occupancy check samples. However, time stamps were missing for 17\% of the meter-level occupancy checks, making those checks unusable. These were distributed roughly evenly across meters, so we had 7 usable samples for most meters.

\subsection{Google Maps Data}

The Google Maps Directions API can be used to estimate travel time between any two points in a city that are connected by roads. Ideally we would simply tell Google Maps the route to drive and report the time to drive it. However, the number of such calls that can be made to the API is limited. Instead, we used the API to find driving and walking times for each block face in the city by getting the travel time between the centers of the intersections at either end of the block. We then reconstructed travel time between pairs of points in the city by labeling the road graph with these values and directly running a shortest path algorithm (e.g., Dijkstra's algorithm).

\subsection{City-Owned Lot Data}

We were given access to proprietary parking data from 21 city-owned lots, containing between 9 and 643 stalls. For each lot we used hourly data from a 3 month period. In each hour the data recorded the number of entries and the amount of time each vehicle paid for. Unfortunately, this did not give us an exact record of when each vehicle left the lot. This problem was particularly pronounced when combined with promotional rates (e.g., a flat rate to park until 6:00 PM). If we naively treated every vehicle as leaving at the moment its parking expired, we would have seen huge artificial spikes at times that represent rate boundaries. 
To deal with this, we applied a Gaussian smoothing procedure to the departure data for each of these peaks. We then redistributed the departures in the peaks back 12 hours, to account for vehicles leaving the lot at different times, with the quantities distributed to each hour proportional to the probabilities of a left-half-Gaussian distribution with mean at the peak and standard deviation 3.5. 

%% file: results.tex
\subsection{Occupancy Prediction Network}\label{sec:nn_results}


To evaluate the performance of the occupancy prediction neural network, we compare it with a logistic regression model as baseline using cross entropy loss on the validation set. Cross entropy is a measure of difference between two probability distributions; a model with low cross entropy loss does a good job of approximating the true distribution of the data. Under this measure, our network on average outperformed the baseline by 9\%: the network achieved average validation cross entropy loss of 0.51 compared to 0.56 for the baseline. We note however that our network was relatively similar to the baseline in terms of validation accuracy, achieving 79\% vs 77\%. In this case we consider cross entropy loss to be a more meaningful measure of performance, because our model's core purpose is to predict a probability (the probability that a vehicle can park in a given block face).

\subsection{Parking Inefficiencies}

\begin{figure*}[t!]
    \centering
    \includegraphics[width=0.8\textwidth]{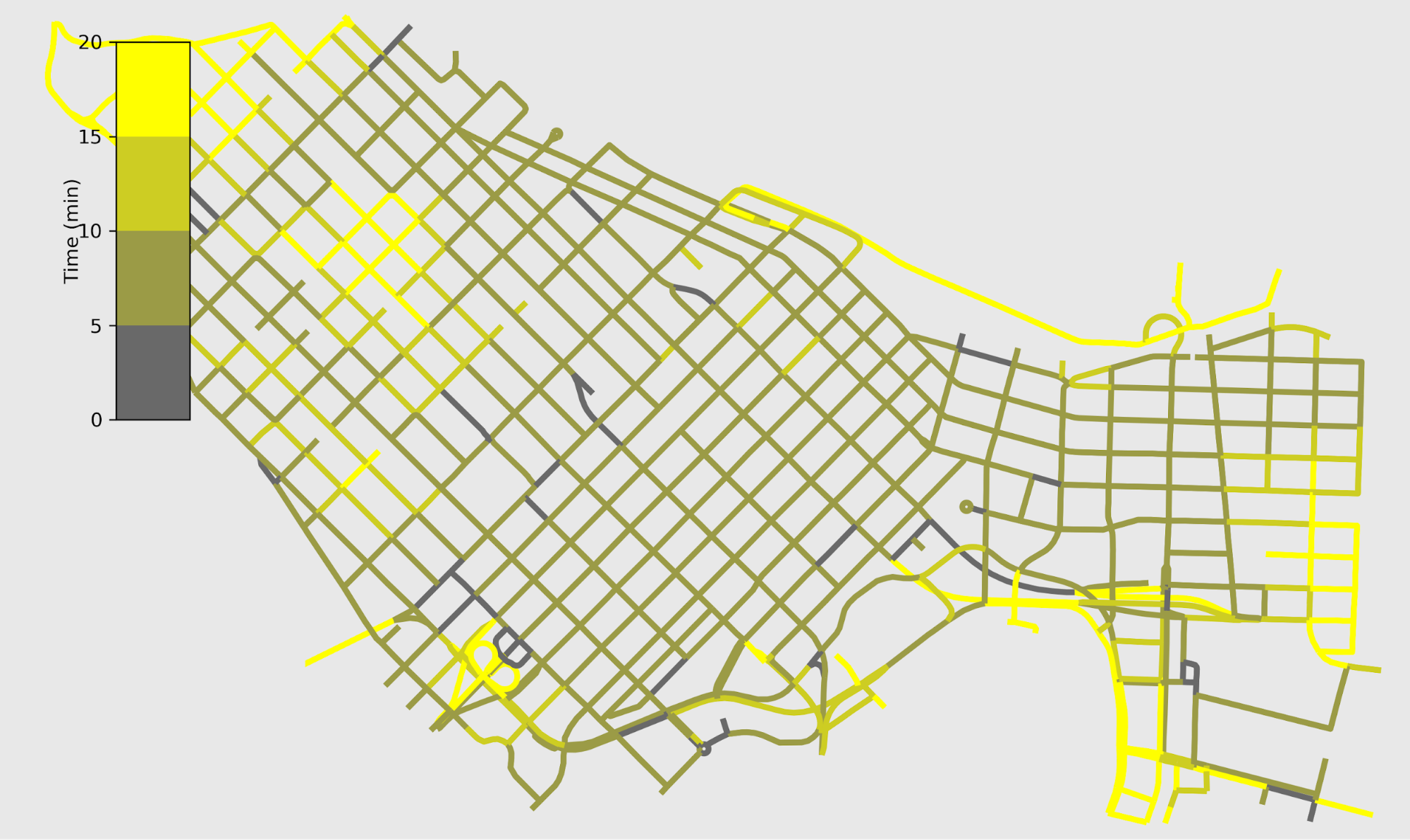}
    \caption{\textit{On Street: Total estimated time to park on-street (including search, park, and walking times) with each block as the assumed destination of the driver. Yellow indicates it will take a relatively long time to park on-street if this block is the driver's final destination. Search times tend to be higher in areas with fewer meters. The map shown is for 12:00-1:00pm on a Friday.} }
    \label{onmap}
\end{figure*}
\begin{figure*}[t!]
    \centering
    \includegraphics[width=0.8\textwidth]{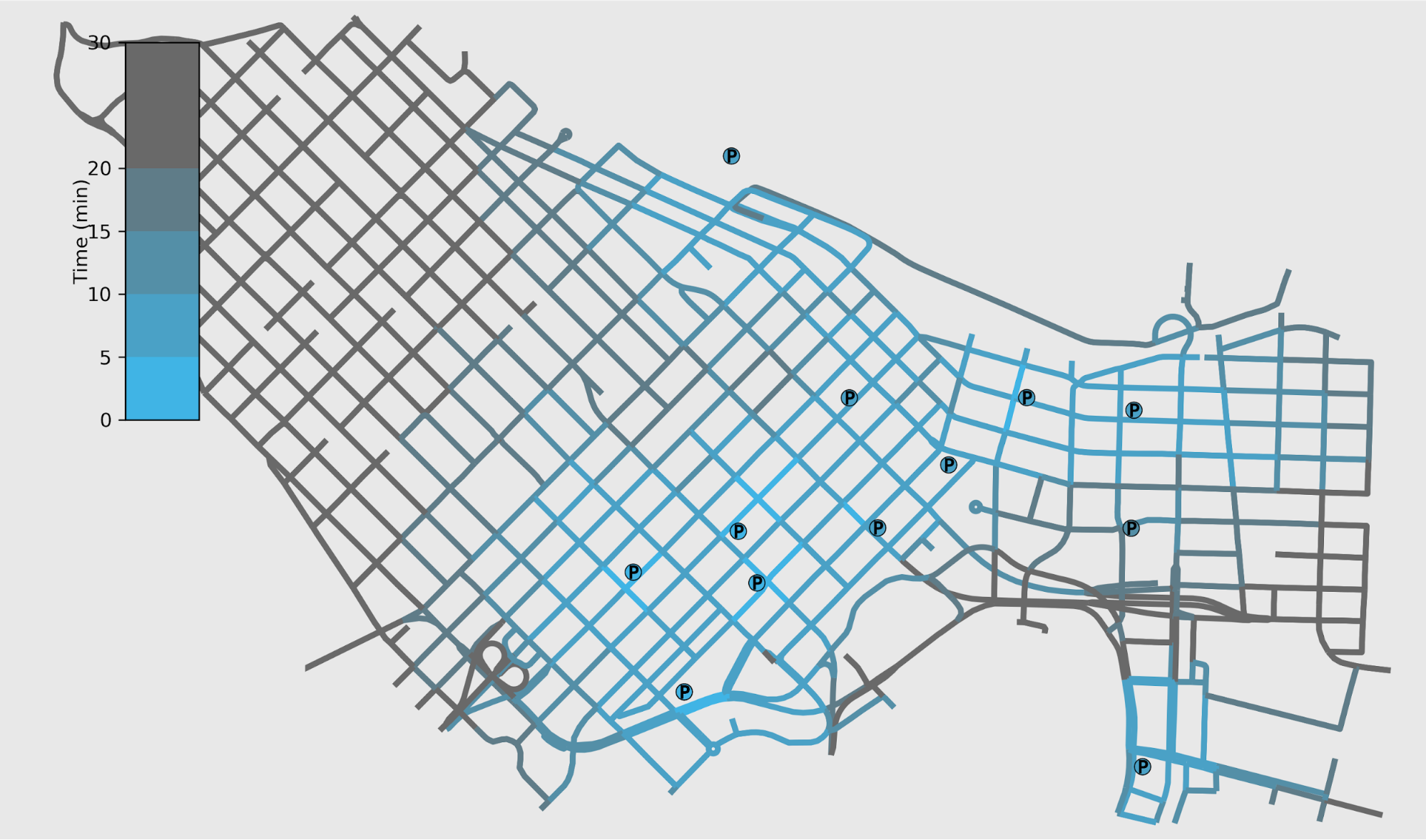}
    \caption{\textit{Off Street: Total estimated time to park in an off-street lot (including driving and walking time, and searching in the lot) with each block as the assumed destination of the driver. Off-street lots are indicated with $P$ icons. Blue indicates a relatively short time to park in a nearby lot if this block is the driver's final destination. Search times tend to be lower in areas nearer to lots. The map shown is for 12:00-1:00pm on a Friday.} }
    \label{offmap}
\end{figure*}

Using the search times generated by our two simulators we estimated the time it would take to park both on- and off-street when driving to a destination in each block in the city. Figures~\ref{onmap} and \ref{offmap} show the total estimated time for parking on-street and off-street, respectively. Not surprisingly, the on-street map (Figure~\ref{onmap}) shows it will take a relatively long time to park on-street in the west end of the city and around the Granville St. area, where meters are less common. Similarly, the off-street map shows that parking off-street is quickest when a driver's destination is near an off-street lot. 

\begin{figure*}[t!]
    \centering
    \includegraphics[width=0.8\textwidth]{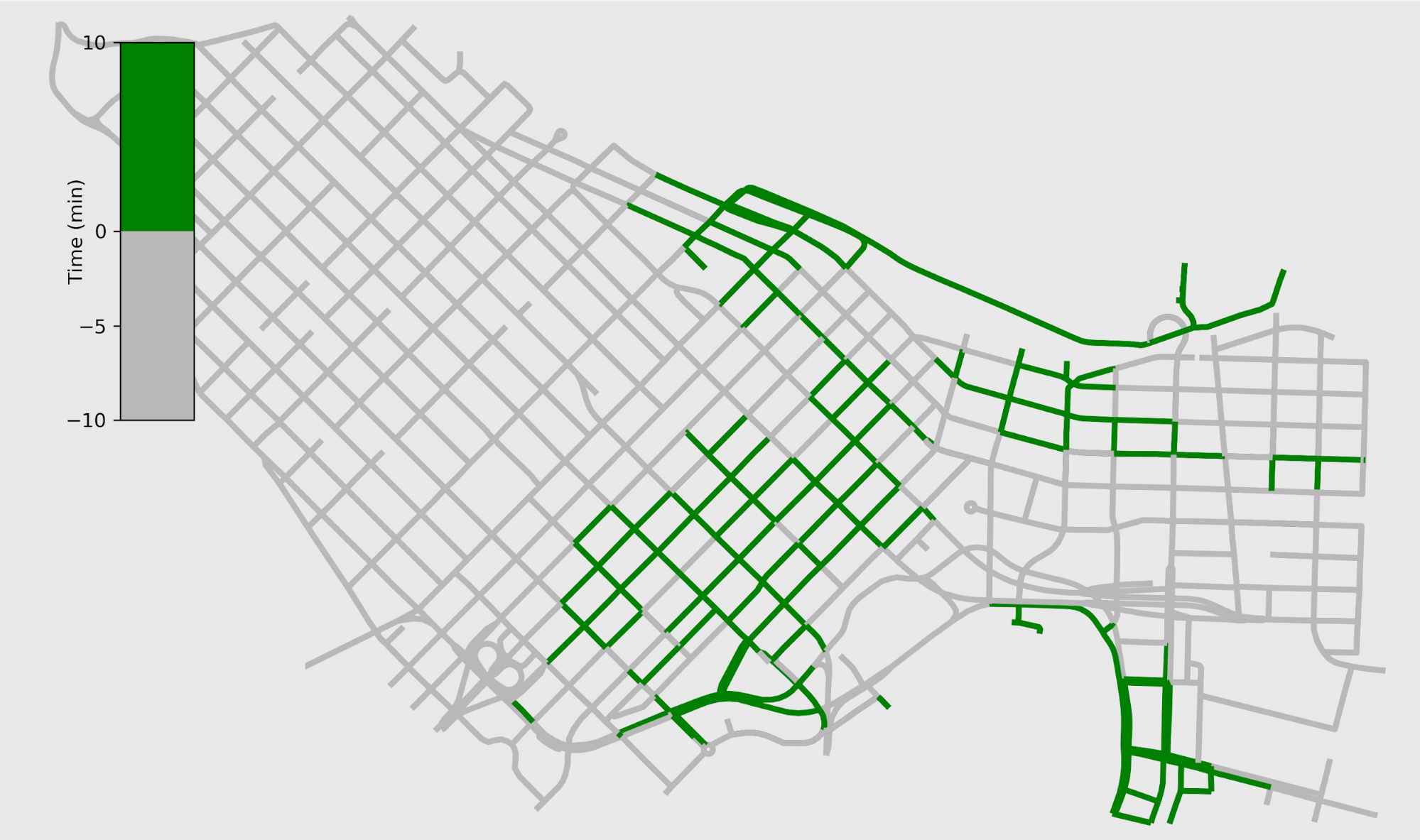}
    \caption{\textit{Areas of Opportunity: Difference in estimated time between parking off-street and parking on-street. Blocks coloured green indicate drivers can actually save time by parking off-street rather than searching for curbside spots if these blocks are their destinations. Curbside parking in these areas can be eliminated, with nearby lots absorbing the excess capacity. This space can then serve more community-friendly purposes. The map shown is for 12:00-1:00pm on a Friday.} }
    \label{diffmap}
\end{figure*}

It is most interesting to examine the difference between these two maps. Figure~\ref{diffmap} shows how much time a driver could \textit{save} by parking off-street, with each block as the assumed destination. Blocks coloured green indicate areas that are at least as accessible from lots as they are from on-street meters. Most of these areas can be reached about 5 minutes \textit{faster} by parking off-street, with some being as much as 10 to 15 minutes faster. In some cases the issue is simply a lack of parking meters on the destination street, which means that drivers must walk some distance regardless of their parking modality. In other areas (e.g., around Granville St., Yaletown, and Chinatown, seen in the center-right of Figure~\ref{diffmap}) congested roads overlap with areas that are easily accessible from off-street lots. In these areas searching for on-street parking can take a long time due to the congestion, but the presence of nearby lots means off-street parking is relatively quick. These areas represent an opportunity for the city to effect change in the way people use parking resources in the city. 

%% file: conclusion.tex
While on-street parking has certain advantages, its overuse has negative side effects that impact everyone in the city. Drivers who are looking for parking occupy valuable road space unnecessarily and also tend to be distracted. Parked cars can obscure pedestrians and cyclists from drivers' sight. At the same time, many off-street parking lots have large amounts of excess capacity at times of day when on-street parking is in high demand. 

We examined parking usage on metered blocks and in city-owned, off-street lots in downtown Vancouver. We used deep learning to predict true on-street occupancy rates from partial payment and other, block-level data. We used Monte Carlo simulation in conjunction with Google Maps driving and walking times to model the amount of time taken to park both on- and off-street, with each block face in the city as the assumed destination. We identified areas where it would be at least as fast for drivers to park in nearby lots as to search the area and then park on-street. 

Since the particular lots we considered are owned by the City, the City has the power to change the way they are operated. Thus, these areas represent a clear opportunity to improve the city as a whole. Curbside space is a valuable resource that modern municipalities can utilize to improve the lives of all their residents. In particular, on-street parking reduces safety for drivers, pedestrians and cyclists alike by obscuring vision and increasing the number of  distracted drivers on the road. What's more, curbside lanes can be converted to dedicated bike or transit lanes, helping the City of Vancouver maintain its vehicle mode-share targets. Alternatively, these lanes can be converted to parklets or green spaces, making these areas more community-friendly. 

How then can drivers be incentivized to use lots instead of curbside space? One simple way is through better information, either in the form of signage or through a website or app showing off street availability. More overt incentives are also feasible. For the most part, off-street parking rates in Vancouver are higher than nearby on-street parking rates. Better aligning pricing and time restrictions between on- and off-street parking could encourage more drivers to move off-street. Most ambitiously, a reservation system could provide the strongest incentives by enabling fixed pricing and eliminating uncertainty for drivers, allowing them to head directly to a lot from their home. Any of these measures could help the City make better use of valuable on-street space at little to no cost to drivers. 